# Super Eightbrane in Superspace[1]

Hitoshi NISHINO

*Department of Physics*
*University of Maryland at College Park*
*College Park, MD 20742-4111, USA*
E-Mail: nishino@nscpmail.physics.umd.edu

## Abstract

We present a superspace formulation for super eightbrane theory based on massive type IIA supergravity in ten-dimensions. Remarkably, in addition to the 10-form superfield strength originally needed for super eightbrane, we also need an 'over-ranked' 11-form superfield strength $H_{A_1 \cdots A_{11}}$ with identically vanishing purely bosonic component, in order to satisfy all the Bianchi identities. As a natural super $p$-brane formulation for the 11-form superfield strength, we present a super ninebrane action on 10-dimensional super-worldvolume invariant under a local fermionic $\kappa$-symmetry. We also show that we can formulate such a superspace with an 'over-ranked' superfield strength also in eleven-dimensions, and possibly in other lower dimensions as well.

---

[1]This work is supported in part by NSF grant # PHY-93-41926.

## 1. Introduction

In 1995, a first possible effective theory formulation [1] for super eightbrane [2] or Dirichlet eightbrane [3] based on *massive* type IIA supergravity [4] was presented. The basic idea is to start with the *massive* type IIA supergravity in ten-dimensions (10D) [4], regarding the mass parameter $m$ as a scalar field, and then perform a duality transformation into its 9-form potential field with 10-form field strength as its Hodge dual. This 9-form potential field is to be identified as the background superfield for the super eightbrane action. However, the drawback of ref. [1] was that only bosonic terms were given with no fermionic partners, which are indispensable as an effective supergravity theory.

Although there appears to be a tremendous amount of support for the idea of the Dirichlet $p$-brane [3], it has been noticed by Gates [5] that there has been no satisfactory superspace [6] (nor component) level formulation accompanying such a 10-form component field strength in a manifestly supersymmetric manner. As a simple consideration reveals, there seems to be a fundamental obstruction for accommodating the 10-form field strength in superspace. To see this, let us introduce the 10-form superfield strength $N_{A_1 \cdots A_{10}}$ corresponding to the 10-form component field strength $N_{a_1 \cdots a_{10}}$ for its 9-form potential $M_{a_1 \cdots a_9}$. The fundamental problem seems to be the contradiction between the non-zero 'constancy' of the bosonic component $N_{a_1 \cdots a_{10}}$, and the satisfaction of the Bianchi identity (BI) for $N_{A_1 \cdots A_{10}}$, that usually occur at the $N$-BI with dimensionality $d = 1$ of the form $\nabla_{(\underline{\alpha}} N_{\underline{\beta})c_1 \cdots c_9} + \cdots \equiv 0$. This is because if $N_{a_1 \cdots a_{10}}$ is constant, then the supersymmetry transformation of the potential should vanish up to gauge transformation: $\delta_Q M_{a_1 \cdots a_9} = 0$. Now due to the general relationship $\delta_Q M_{c_1 \cdots c_9} = \epsilon^{\underline{\alpha}} N_{\underline{\alpha} c_1 \cdots c_9} + (1/8!) \psi_{[c_1|}{}^{\underline{\beta}} \epsilon^{\underline{\alpha}} N_{\underline{\alpha}\underline{\beta}|c_2 \cdots c_9]}$ for a supersymmetry transformation $\delta_Q$ [6] we expect no such components as $N_{\underline{\alpha}\underline{\beta}c_1 \cdots c_8}$. However, once this component is zero, there seems no way to satisfy the above $N$-BI in superspace, keeping $N_{a_1 \cdots a_{10}}$ as a non-zero constant.

In this paper, we will finally overcome this difficulty by introducing an over-ranked 11-form superfield strength $H_{A_1 \cdots A_{11}}$, mixed up with the $N$-BI. Such over-ranked superfield strengths have been already considered in 1980 [7], but our 11-form superfield strength is more unconventional, because it enters into another superfield strength $N_{A_1 \cdots A_{10}}$ *via* a 'generalized Chern-Simons term', as will be seen. We will also give a possible super ninebrane action coupled to the 10-form potential superfield *via* a Wess-Zumino-Novikov-Witten (WZNW) term. As a by-product, we provide a superspace formulation with a 12-form superfield strength for 11D supergravity [8][9]. Finally, we also give conditions for any possible supergravity theory with an over-ranked $(D+1)$-form superfield strength in $D$-dimensional space-time.

## 2. Superspace Constraints

We start with listing up our BIs that are crucial for our formulation. First of all, our component field content is $(e_m{}^a, \psi_m{}^{\underline{\alpha}}, A_m, B_{mn}, A_{mnp}, \Phi, M_{m_1\cdots m_9}, C_{m_1\cdots m_{10}})$,[2] where the first six fields are the standard ones for type IIA supergravity [10] or its massive generalization [4][11], e.g., $A_m$, $B_{mn}$ and $A_{mnp}$ respectively have the field strengths $F_{mn}$, $G_{mnp}$ and $F_{mnpq}$, while $M_{m_1\cdots m_9}$ and $C_{m_1\cdots m_{10}}$ are new in our superspace formulation. In particular, $M_{m_1\cdots m_9}$ has the field strength $N_{m_1\cdots m_{10}}$, while $C_{m_1\cdots m_{10}}$ has an identically vanishing field strength $H_{m_1\cdots m_{11}} \equiv 0$ obviously in 10D. Accordingly, our superfield strengths are $T_{AB}{}^C$, $R_{ABc}{}^d$, $F_{AB}$, $G_{ABC}$, $F_{ABCD}$, $N_{A_1\cdots A_{10}}$, $H_{A_1\cdots A_{11}}$, where the first three are the standard ones in type IIA supergravity [10][12], or its massive generalization [4][11], while the last two are peculiar to our superspace formulation for super eightbrane. Especially, despite of $H_{a_1\cdots a_{11}} \equiv 0$, there are non-vanishing components $H_{\underline{\alpha}\underline{\beta}c_1\cdots c_9}$, while $N_{a_1\cdots a_{10}}$ is the only non-zero component among $N_{A_1\cdots A_{10}}$, as will be seen in (2.10).

All of our superspace BIs to be confirmed are listed up, as

$$\tfrac{1}{2}\nabla_{[A}F_{BC)} - \tfrac{1}{2}T_{[AB|}{}^D F_{D|C)} \equiv 0 ~, \tag{2.1}$$

$$\tfrac{1}{6}\nabla_{[A}G_{BCD)} - \tfrac{1}{4}T_{[AB|}{}^E G_{E|CD)} \equiv 0 ~, \tag{2.2}$$

$$\tfrac{1}{24}\nabla_{[A}F_{BCDE)} - \tfrac{1}{12}T_{[AB|}{}^F F_{F|CDE)} - \tfrac{1}{12}F_{[AB}G_{CDE)} \equiv 0 ~, \tag{2.3}$$

$$\tfrac{1}{10!}\nabla_{[A_1}N_{A_2\cdots A_{11})} - \tfrac{1}{2(9!)}T_{[A_1 A_2|}{}^B N_{B|A_3\cdots A_{11})} - H_{A_1\cdots A_{11}}N \equiv 0 ~. \tag{2.4}$$

$$\tfrac{1}{11!}\nabla_{[A_1}H_{A_2\cdots A_{12})} - \tfrac{1}{2(10!)}T_{[A_1 A_2|}{}^B H_{B|A_3\cdots A_{12})} \equiv 0 ~, \tag{2.5}$$

$$\tfrac{1}{2}\nabla_{[A}T_{BC)}{}^D - \tfrac{1}{2}T_{[AB|}{}^E T_{E|C)}{}^D - \tfrac{1}{4}R_{[AB|f}{}^g\left(\mathcal{M}_g{}^f\right)_{|C)}{}^D \equiv 0 ~. \tag{2.6}$$

Our (anti)symmetrization is defined by $P_{[A}Q_{B)} \equiv P_A Q_B \mp P_B Q_A$, with *no* normalization. The 'constant' scalar (0-form) superfield $N$ in (2.4) is *defined* by

$$N \equiv +\tfrac{1}{10!}\epsilon^{a_1\cdots a_{10}} N_{a_1\cdots a_{10}} ~, \tag{2.7}$$

as the Hodge dual to the 10-form field strength $N_{a_1\cdots a_{10}}$. The constancy $N = \text{const.} \equiv m \neq 0$ in superspace is equivalent to

$$\nabla_A N \equiv 0 ~. \tag{2.8}$$

The BIs (2.4) and (2.5) are our new BIs in this paper, and in particular, the last $HN$-term in (2.4) is the most crucial term, regarded as a result of a generalized Chern-Simons term in $N_{A_1\cdots A_{10}}$, solving the long-standing problem with the $N$-BI. To be more explicit, our $N_{A_1\cdots A_{10}}$ should be defined by

$$N_{A_1\cdots A_{10}} \equiv \tfrac{1}{9!}\nabla_{[A_1} M_{A_2\cdots A_{10})} - \tfrac{1}{2(8!)}T_{[A_1 A_2|}{}^B M_{B|A_3\cdots A_{10})} + C_{A_1\cdots A_{10}} N \quad , \tag{2.9}$$

to comply with (2.4). Here the last $CN$-term can be interpreted as a generalized Chern-Simons term, because it is a product of a potential superfield $C_{A_1\cdots A_{10}}$ and the superfield strength $N$. Also to be stressed is the importance of the constancy (2.8), which makes the whole system work consistently.

---

[2] We use $m, n, \cdots = 0, 1, \cdots, 9$ for 10D curved coordinates, while $a, b, \cdots = (0), (1), \cdots, (9)$ for local Lorentz coordinates in 10D. We sometimes use the *underlined* spinorial indices to symbolize both dotted and undotted components collectively: $\underline{\alpha} \equiv (\alpha, \dot{\alpha})$, $\underline{\beta} \equiv (\beta, \dot{\beta})$, $\cdots$, where $\alpha, \beta, \cdots = 1, 2, \cdots, 16$, $\dot{\alpha}, \dot{\beta}, \cdots = \dot{1}, \dot{2}, \cdots, \dot{16}$. The superspace indices $A, B, \cdots = (a, \underline{\alpha}), (b, \underline{\beta}), \cdots$ are for local Lorentz coordinates.

Even though our BIs above look so simple, we emphasize that we have tried many other options, such as introducing other superfield strengths, such as $F_{A_1\cdots A_8}$ and/or $F_{A_1\cdots A_6}$ dual to $F_{ABCD}$, like the M-5-brane formulations [13][14]. Even though we have also allowed possible Chern-Simons terms for each of these BIs, these superfield strengths never helped us to solve the problem with the unwanted terms in the $N$-BI at $d=1$, as has been mentioned in the Introduction. It seems that the introduction of the 'over-ranked' $H_{A_1\cdots A_{11}}$ is the only solution to this problem in superspace formulation, once $N_{A_1\cdots A_{10}}$ dual to a scalar superfield $N = \text{const.}$ is introduced.

As is well-known, there can be infinitely many sets of constraints in superspace satisfying our BIs. However, we choose in this paper the simplest set, which is sometimes called 'beta-function favored constraints' (BFFC), drastically simplifying the $\beta$-function computation, originally for $N=1$ supergravity in 10D [15], and applied later to type IIA supergravity in [12][3]. Our result for superspace constraints is summarized as

$$T_{\alpha\beta}{}^c = +i(\sigma^c)_{\alpha\beta} \quad , \qquad T_{\dot\alpha\dot\beta}{}^c = +i(\sigma^c)_{\dot\alpha\dot\beta} \quad , \tag{2.10a}$$

$$T_{\alpha\beta}{}^\gamma = +\delta_{(\alpha}{}^\gamma \chi_{\beta)} + (\sigma^c)_{\alpha\beta}(\sigma_c\chi)^\gamma \quad , \qquad T_{\dot\alpha\dot\beta}{}^{\dot\gamma} = +\delta_{(\dot\alpha}{}^{\dot\gamma}\chi_{\dot\beta)} + (\sigma^c)_{\dot\alpha\dot\beta}(\sigma_c\chi)^{\dot\gamma} \quad , \tag{2.10b}$$

$$T_{ab}{}^\gamma = -\tfrac{1}{8}(\sigma^{cd})_\alpha{}^\gamma G_{bcd} \quad , \qquad T_{\dot\alpha b}{}^{\dot\gamma} = +\tfrac{1}{8}(\sigma^{cd})_{\dot\alpha}{}^{\dot\gamma} G_{bcd} \quad , \tag{2.10c}$$

$$T_{ab}{}^{\dot\gamma} = +\tfrac{i}{16}(\sigma_b\sigma^{cd})_\alpha{}^{\dot\gamma}\left(\epsilon^{-\Phi}F_{cd} + \chi_{cd}\right) - \tfrac{i}{8}(\sigma_b)_\alpha{}^{\dot\gamma}(\chi\chi)$$
$$\quad + \tfrac{i}{192}(\sigma_b\sigma^{[4]})_\alpha{}^{\dot\gamma}\left(e^{-\Phi}F_{[4]} - \chi_{[4]}\right) + \tfrac{i}{8}(\sigma_b)_\alpha{}^{\dot\gamma} e^{-\Phi} N \quad , \tag{2.10d}$$

$$T_{\dot\alpha b}{}^\gamma = -\tfrac{i}{16}(\sigma_b\sigma^{cd})_{\dot\alpha}{}^\gamma\left(\epsilon^{-\Phi}F_{cd} + \chi_{cd}\right) - \tfrac{i}{8}(\sigma_b)_{\dot\alpha}{}^\gamma(\chi\chi)$$
$$\quad + \tfrac{i}{192}(\sigma_b\sigma^{[4]})_{\dot\alpha}{}^\gamma\left(e^{-\Phi}F_{[4]} - \chi_{[4]}\right) + \tfrac{i}{8}(\sigma_b)_{\dot\alpha}{}^\gamma e^{-\Phi} N \quad , \tag{2.10e}$$

$$F_{\alpha\dot\beta} = +C_{\alpha\dot\beta}\, e^\Phi - N B_{\alpha\dot\beta} \quad , \qquad F_{\alpha\beta} = -N B_{\alpha\beta} \quad , \qquad F_{\dot\alpha\dot\beta} = -N B_{\dot\alpha\dot\beta} \quad , \tag{2.10f}$$

---
[3]Even though the basic structure for BFFC in [12] is valid, there are some errors in numerical coefficients in the superspace constraints given there. See below (2.13).

$$F_{\alpha b} = +ie^\Phi(\sigma_b\chi)_\alpha - NB_{\alpha b} \;, \qquad F_{\dot{\alpha} b} = +ie^\Phi(\sigma_b\chi)_{\dot{\alpha}} - NB_{\dot{\alpha} b} \;, \tag{2.10g}$$

$$G_{\alpha\beta c} = +i(\sigma^c)_{\alpha\beta} \;, \qquad G_{\dot{\alpha}\dot{\beta} c} = -i(\sigma^c)_{\dot{\alpha}\dot{\beta}} \;, \tag{2.10h}$$

$$F_{\alpha\dot{\beta} cd} = +e^\Phi(\sigma_{cd})_{\alpha\dot{\beta}} + Y_{\alpha\dot{\beta} cd} \;, \tag{2.10i}$$

$$F_{\alpha bcd} = +ie^\Phi(\sigma_b\chi)_\alpha + Y_{\alpha bcd} \;, \qquad F_{\dot{\alpha} bcd} = +ie^\Phi(\sigma_b\chi)_{\dot{\alpha}} + Y_{\dot{\alpha} bcd} \;, \tag{2.10j}$$

$$\nabla_\alpha \Phi = +\chi_\alpha \;, \qquad \nabla_{\dot{\alpha}} \Phi = +\chi_{\dot{\alpha}} \;, \qquad \nabla_\alpha \chi_{\dot{\beta}} = -\nabla_{\dot{\beta}} \chi_\alpha \;, \tag{2.10k}$$

$$\nabla_\alpha \chi_\beta = +\tfrac{i}{2}(\sigma^c)_{\alpha\beta} \nabla_c \Phi - \tfrac{i}{24}(\sigma^{[3]})_{\alpha\beta} G_{[3]} - \chi_\alpha \chi_\beta \;, \tag{2.10$\ell$}$$

$$\nabla_{\dot{\alpha}} \chi_{\dot{\beta}} = +\tfrac{i}{2}(\sigma^c)_{\dot{\alpha}\dot{\beta}} \nabla_c \Phi + \tfrac{i}{24}(\sigma^{[3]})_{\dot{\alpha}\dot{\beta}} G_{[3]} - \chi_{\dot{\alpha}} \chi_{\dot{\beta}} \;, \tag{2.10m}$$

$$\nabla_\alpha \chi_{\dot{\beta}} = -\tfrac{3}{16}(\sigma^{cd})_{\alpha\dot{\beta}}(e^{-\Phi} F_{cd} + \chi_{cd}) + \tfrac{5}{8} C_{\alpha\dot{\beta}} (\chi\chi)$$
$$\qquad - \tfrac{1}{192}(\sigma^{[4]})_{\alpha\dot{\beta}}(e^{-\Phi} F_{[4]} - \chi_{[4]}) - \tfrac{5}{8} C_{\alpha\dot{\beta}} e^{-\Phi} N \;, \tag{2.10n}$$

$$H_{\alpha\beta c_1\cdots c_9} = +i(\sigma_{c_1\cdots c_9})_{\alpha\beta} \;, \qquad H_{\dot{\alpha}\dot{\beta} c_1\cdots c_9} = -i(\sigma_{c_1\cdots c_9})_{\dot{\alpha}\dot{\beta}} \;. \tag{2.10p}$$

Here the symbols such as $_{[4]}$ denote the total anti-symmetrizations of bosonic indices, e.g., $A_{[4]} B^{[4]} \equiv A_{abcd} B^{abcd}$ for the totally antisymmetric tensors $A_{abcd}$ and $B^{abcd}$. Other notations are the same as those in ref. [16], e.g., our 10D metric is $(\eta_{ab}) = \text{diag.}(+,-,-,\cdots,-)$ and the $\epsilon$-tensor and $\sigma_{11}$-matrix are defined by $\epsilon^{01\cdots 9} = +1$, $\sigma_{11} \equiv \sigma_{01\cdots 9}$, so that $\sigma_{[10-n]} = (1/n!)(-1)^{[(n-1)/2]} \epsilon_{[10-n]}{}^{[n]} \sigma_{[n]} \sigma_{11}$, and $(1/n!) \epsilon_{a_1\cdots a_{10-n}}{}^{[n]} \epsilon_{[n]}{}^{c_1\cdots c_{10-n}} = -(-1)^{[(n-1)/2]} \times \delta_{[a_1}{}^{c_1} \cdots \delta_{a_{10-n}]}{}^{c_{10-n}}$, where $[(n-1)/2]$ in exponents is the Gauß's symbol for the integer part of $(n-1)/2$. Our fermionic index contraction rule is the same as that in [16], i.e., the contraction between northwest and southeast is the standard one with no extra sign, while that between southwest and northeast costs an extra sign, e.g., $(\sigma_c \chi)^\alpha \equiv (\sigma_c)^{\alpha\beta} \chi_\beta \equiv -(\sigma_c)^\alpha{}_{\dot{\beta}} \chi^{\dot{\beta}}$ also with flippings of the dottedness of indices, when raised and lowered by the charge conjugation matrix $C^{\alpha\beta} = -C^{\beta\alpha}$, $C_{\alpha\dot{\beta}} = -C_{\dot{\beta}\alpha}$, e.g., $(\sigma_c \chi)^\alpha = +C^{\alpha\dot{\beta}}(\sigma_c \chi)_{\dot{\beta}} = -(\sigma_c \chi)_{\dot{\beta}} C^{\dot{\beta}\alpha}$, etc.[4] The symbols $\chi_{[2]}$, $\chi_{[4]}$ and $(\chi\chi)$ are for

$$\chi_{[2]} \equiv \chi^\alpha (\sigma_{[2]})_\alpha{}^\beta \chi_\beta \;, \qquad \chi_{[4]} \equiv \chi^\alpha (\sigma_{[4]})_\alpha{}^\beta \chi_\beta \;, \qquad (\chi\chi) \equiv \chi^\alpha \chi_\alpha \;. \tag{2.11}$$

Note that all the other remaining independent components for the superfields in (2.10), such as $T_{\underline{a}b}{}^c$ are all zero, in particular, $N_{\underline{\alpha} b_1\cdots b_9} = 0$, $N_{\underline{\alpha}\beta c_1\cdots c_9} = 0$, $H_{\underline{\alpha} c_1\cdots c_{10}} = 0$, $H_{c_1\cdots c_{11}} \equiv 0$. The $Y_{ABCD}$ is the super Chern-Simons form defined by [11]

$$Y_{ABCD} \equiv \tfrac{1}{4} F_{[AB} B_{CD)} \;. \tag{2.12}$$

---

[4] Just as a guide for the readers unfamiliar with this notation, other important relations are such as $(\sigma^{[2]})_{\dot{\beta}\alpha} \equiv +(\sigma^{[2]})_{\alpha\dot{\beta}}$, $(\sigma^{[4]})_{\dot{\beta}\alpha} \equiv -(\sigma^{[4]})_{\alpha\dot{\beta}}$, $(\sigma^{[2]})_\alpha{}^\beta = +(\sigma^{[2]})^\beta{}_\alpha$, $(\sigma^{[4]})_\alpha{}^\beta = -(\sigma^{[4]})^\beta{}_\alpha$, etc.

Note also that $F_{mn}$ is defined by $F_{mn} \equiv \partial_{[m} A_{n]} + N B_{mn}$ as usual for a massive type IIA formulation [4][11] for the mass $m = N$, complying with (2.10f) and (2.10g).

We now briefly describe some crucial points for the confirmation of BIs at $d \leq 1$. The most crucial step is the introduction of the $HN$-term in (2.4). Aa was already mentioned in the Introduction, if there were no such a term in (2.4), there would be no satisfaction of the $N$-BI for the component $(\underline{\alpha\beta} c_1 \cdots c_9)$ at $d = 1$. There are other simpler and less crucial confirmations, such as the $(\alpha\beta c_1 \cdots c_{10})$-component of the $H$-BI (2.5), which has only one term $(\sigma_{[c_1\cdots c_8|}{}^d)_{\alpha\beta} G_{|c_9 c_{10}]d}$ which does not seem to vanish by itself. However, multiplication of this term by $\epsilon^{c_1\cdots c_{10}}$ immediately reveals its vanishing, due to

$$\epsilon^{c_1\cdots c_{10}} \epsilon_{c_1\cdots c_8}{}^{de} (\sigma_e)_{\alpha\beta} G_{c_9 c_{10} d} \equiv 0 \ . \tag{2.13}$$

This is because the index $d$ on $G_{c_9 c_{10} d}$ should be only $c_9$ or $c_{10}$, yielding zero for the totally antisymmetric $G_{abc}$.[5]

We mention some errors detected in various numerical coefficients in ref. [12], which are now corrected by our constraints above. The most crucial ones in [12] are for the $F_{[4]}$-term in $T_{ab}{}^{\dot\gamma}$, $T_{\dot\alpha b}{}^{\gamma}$, and $\nabla_\alpha \chi_{\dot\beta}$. Similar errors are found in the coefficients for $\chi^2$-terms, in particular, the $(\chi\chi)$-term is missing in $T_{ab}{}^{\dot\gamma}$ in [12]. Even though the $\chi^2$-terms are not essential at the lowest order, they will affect many computations at the bilinear order terms with a lot of inconsistency, such as $\chi N$-terms described in the next section.

## 3. BIs at $d = 3/2$ and Fermionic Superfield Equations

We can perform a simple consistency check about our superspace formulation with the modifications with the $N$-superfield strength accompanied by the additional superfield strength $H_{A_1\cdots A_{11}}$ at $d = 3/2$. The most crucial test is the $(\underline{\alpha} c_1 \cdots c_{10})$-type $N$-BI, which is shown to hold *only* under $\nabla_A N = 0$. Another important confirmation is the fermionic superfield equations, obtained from the $(\alpha\beta c, \gamma)$-type $T$-BI (2.6) at $d = 3/2$. Our fermionic superfield equations with the $N$-modifications thus obtained are

$$i(\sigma^b T_{ab})_\alpha + 2\nabla_a \chi_\alpha = (\chi G \text{ and } \chi\nabla\Phi\text{-terms}) \ , \tag{3.1a}$$

$$(\sigma^{ab} T_{ab})_{\dot\alpha} = +\tfrac{1}{2}\chi_{\dot\alpha} N + (\chi G \text{ and } \chi\nabla\Phi\text{-terms}) \ , \tag{3.1b}$$

$$2i(\slashed{\nabla}\chi)_{\dot\alpha} = +\tfrac{1}{4}\chi_{\dot\alpha} N + (\chi G \text{ and } \chi\nabla\Phi\text{-terms}) \ , \tag{3.1c}$$

and similar forms for the other chiral components. Note that there is no $\chi N$-term in (3.1a). Since $N = \text{const.} = m \neq 0$, the $N$-dependent terms in (3.1) are regarded as 'mass' terms, corresponding to the original component formulation [4].

---
[5]Similar identities for higher-rank tensors have been already mentioned in ref. [17]

There are several ways of getting these fermionic superfield equations, and they yield consistent results. Let $X_{\underline{\alpha}\underline{\beta}c}{}^{\underline{\delta}}$ be the l.h.s. of $(\underline{\alpha}\underline{\beta}c,\underline{\delta})$-component of the $T$-BI of (2.6). We have performed the contractions (i) $\delta_{\underline{\delta}}{}^{\underline{\beta}} X_{\underline{\alpha}\underline{\beta}c}{}^{\underline{\delta}} \equiv 0$, (ii) $i(\sigma^c)^{\underline{\alpha}\underline{\beta}} X_{\underline{\alpha}\underline{\beta}c}{}^{\underline{\delta}} \equiv 0$, (iii) $C^{\underline{\alpha}\underline{\dot\beta}} X_{\underline{\alpha}\underline{\dot\beta}c}{}^{\underline{\delta}} \equiv 0$, (iv) $\delta_{\underline{\dot\delta}}{}^{\underline{\alpha}} X_{\underline{\alpha}\underline{\dot\beta}c}{}^{\underline{\delta}} \equiv 0$. The contraction (i) gives directly (3.1a), while (ii), (iii) and (iv) give the consistent equations (3.1b) and (3.1c). This is one of the non-trivial consistency check in our formulation, in particular with the new 11-form superfield strength. As a matter of fact, there occur lots of highly non-trivial cancellations among all the unwanted terms, providing good cross-checks for the whole system. Relevantly, the $\chi N$-terms in (3.1) are sensitive to the numerical coefficients of $F_{[4]}$ and $\chi_{[4]}$-terms in the constraints (2.10d), (2.10e) and (2.10n). In fact, this led us to detect the errors in numerical coefficients given in [12].

## 4. Super Ninebrane in 10D

The existence of the 11-form superfield strength $H_{A_1\cdots A_{11}}$ strongly indicates the possible super *ninebrane* formulation on 10D world-supervolume. We give here a brief description of such a formulation, as a special case of more general super $p$-brane formulations [2].[6]

Our ansatz for the total action of super ninebrane is

$$I_{10} \equiv \int d^{10}\sigma \left[ +\tfrac{1}{2}\sqrt{-g}\, g^{ij}\Pi_i{}^a\Pi_{ja} - 4\sqrt{-g} - \tfrac{1}{10!}\epsilon^{i_1\cdots i_{10}}\Pi_{i_1}{}^{A_1}\cdots \Pi_{i_{10}}{}^{A_{10}} C_{A_{10}\cdots A_1} \right], \quad (4.1)$$

with the 10-form potential superfield $C_{A_1\cdots A_{10}}$ for the 11-form superfield strength $H_{A_1\cdots A_{11}}$. The indices $i, j, \cdots = 0, 1, \cdots, 9$ are for the curved 10D world-supervolume coordinates with the metric $g_{ij}$. The $\Pi_i{}^A \equiv (\partial_i Z^M) E_M{}^A$ with the target space-time (inverse) vielbein $E_M{}^A$ is the usual pull-back in super $p$-brane formulation [2]. Our action $I_{10}$ is invariant under the fermionic $\kappa$-symmetry [2]:

$$\delta_\kappa E^{\underline{\alpha}} = \tfrac{1}{2}(I+\Gamma)^{\underline{\alpha}}{}_{\underline{\beta}} \kappa^{\underline{\beta}}, \qquad \delta_\kappa E^a = 0, \quad (4.2a)$$

$$\Gamma_{\underline{\alpha}}{}^{\underline{\beta}} \equiv \tfrac{1}{10!\sqrt{-g}} \epsilon^{i_1\cdots i_{10}} \Pi_{i_1}{}^{a_1}\cdots \Pi_{i_{10}}{}^{a_{10}} (\sigma_{a_{10}\cdots a_1})_{\underline{\alpha}}{}^{\underline{\beta}}. \quad (4.2b)$$

Here $\delta_\kappa E^A \equiv (\delta_\kappa Z^M) E_M{}^A$, as usual [2], and the matrix $I+\Gamma$ plays a role of a projection operator, and it is easy to show that

$$\Gamma^2 = +I. \quad (4.3)$$

Note that the dimension of our world-supervolume and that of the target 10D space-time coincide. This further implies that $\Gamma$ defined by (4.2b) is equal to the $\sigma_{11}$-matrix for our 10D superspace:

$$\Gamma = \sigma_{11}. \quad (4.4)$$

---
[6]Due to the lack of strong motivation, the $(p+1)$-dimensional target space-time for super $p$-brane was ignored in Table 1 in ref. [2].

The proof of this relation relies also on the embedding condition

$$g_{ij} = \Pi_i{}^a \Pi_{ja} ~, \tag{4.5}$$

obtained as the algebraic field equation of $g_{ij}$. Notice that $\Pi_i{}^a$ can be identified with the world-supervolume zehnbein up to appropriate Lorentz transformations, because of the matching range of indices of both $i$ and $a$, namely $-\det(g_{ij}) = (\det \Pi_i{}^a)^2$. Using this, we can confirm (4.4) as

$$\begin{aligned}(\text{LHS of (4.4)}) &= -\tfrac{1}{10!\sqrt{-g}} \epsilon^{i_1 \cdots i_{10}} \Pi_{i_1}{}^{a_1} \cdots \Pi_{i_{10}}{}^{a_{10}} (\epsilon_{a_{10} \cdots a_1} \sigma_{11}) \\ &= +\tfrac{1}{10!\sqrt{-g}} \epsilon^{i_1 \cdots i_{10}} \epsilon_{a_1 \cdots a_{10}} \Pi_{i_1}{}^{a_1} \cdots \Pi_{i_{10}}{}^{a_{10}} \sigma_{11} \\ &= \tfrac{1}{\sqrt{-g}} (\det \Pi_i{}^a)(\sigma_{11}) = +\sigma_{11} = (\text{RHS of (4.4)}) ~. \end{aligned} \tag{4.6}$$

In order to simplify other computations, it is convenient to use the notation

$$(\Gamma_{i_1 \cdots i_n})_{\underline{\alpha\beta}} \equiv \Pi_{i_1}{}^{a_1} \cdots \Pi_{i_n}{}^{a_n} (\sigma_{a_1 \cdots a_n})_{\underline{\alpha\beta}} ~. \tag{4.7}$$

The $\Gamma_i$ matrices conveniently satisfy the Clifford algebra $\{\Gamma_i, \Gamma_j\} = +2g_{ij}$. Armed with these relations, we can now easily verify the $\kappa$-invariance of our action (4.1), by the aid of other relations such as

$$\epsilon^{i_1 \cdots i_{10}} \Gamma_{i_9 \cdots i_1} = +(9!) \sqrt{-g} \Gamma^{i_1} \Gamma ~, \tag{4.8}$$

which seems by now almost trivial. In fact, an intermediate stage of the variation of $I_{10}$ looks like

$$\begin{aligned}\delta_\kappa I_{10} =~ &-i\sqrt{-g}\,\Pi_i{}^{\underline{\gamma}} (\Gamma^i)_{\underline{\gamma\beta}} \tfrac{1}{2}(I+\Gamma)^{\underline{\beta}}{}_{\underline{\delta}} \kappa^{\underline{\delta}} \\ &- \tfrac{1}{9!} \epsilon^{i_1 \cdots i_{10}} \tfrac{1}{2}(I+\Gamma)^{\underline{\beta}}{}_{\underline{\gamma}} \kappa^{\underline{\gamma}} i (\Gamma_{i_9 \cdots i_1} \Gamma)_{\underline{\beta\alpha}} \Pi_{i_{10}}{}^{\underline{\alpha}} ~,\end{aligned} \tag{4.9}$$

which vanishes under the relations above.

### 5.  12-Form Superfield Strength in 11D

Once we have understood this super ninebrane formulation in 10D with the peculiar 11-form superfield strength, our natural question is whether such a feature is common to other superspace formulations in other dimensions. The answer to this question seems affirmative, and we first give an explicit example for 11D supergravity [9][8].

Mimicking our 10D result, we introduce two extra superfield strengths $N_{A_1 \cdots A_{11}}$ and $H_{A_1 \cdots A_{12}}$, in addition to the conventional ones $T_{AB}{}^C$, $R_{ABc}{}^d$ and $F_{ABCD}$ in 11D [8]. Now

our BIs are

$$\tfrac{1}{24}\nabla_{[A}F_{BCDE)} - \tfrac{1}{12}T_{[AB|}{}^F F_{F|CDE)} \equiv 0 ~, \tag{5.1}$$

$$\tfrac{1}{11!}\nabla_{[A_1}N_{A_2\cdots A_{12})} - \tfrac{1}{2(10!)}T_{[A_1A_2|}{}^B N_{B|A_3\cdots A_{12})} + H_{A_1\cdots A_{12}}N \equiv 0 ~, \tag{5.2}$$

$$\tfrac{1}{12!}\nabla_{[A_1}H_{A_2\cdots A_{13})} - \tfrac{1}{2(11!)}T_{[A_1A_2|}{}^B H_{B|A_3\cdots A_{13})} \equiv 0 ~, \tag{5.3}$$

$$\tfrac{1}{2}\nabla_{[A}T_{BC)}{}^D - \tfrac{1}{2}T_{[AB|}{}^E T_{E|C)}{}^D - \tfrac{1}{4}R_{[AB|e}{}^f (\mathcal{M}_f{}^e)_{|C)}{}^D \equiv 0 ~. \tag{5.4}$$

The $HN$-term in (5.2) is an 11D analog of that in (2.4).

Our superspace constraints are

$$T_{\alpha\beta}{}^c = +i(\gamma^c)_{\alpha\beta} ~, \qquad F_{\alpha\beta cd} = +\tfrac{1}{2}(\gamma_{cd})_{\alpha\beta} ~, \tag{5.5a}$$

$$T_{ab}{}^\gamma = +\tfrac{i}{144}(\gamma_b{}^{[4]}F_{[4]} + 8\gamma^{[3]}F_{b[3]})_\alpha{}^\gamma ~, \tag{5.5b}$$

$$N \equiv +\tfrac{1}{11!}\epsilon^{a_1\cdots a_{11}} N_{a_1\cdots a_{11}} = \text{const.} ~, \qquad \nabla_A N = 0 ~, \tag{5.5c}$$

$$H_{\alpha\beta c_1\cdots c_{10}} = +(\gamma_{c_1\cdots c_{10}})_{\alpha\beta} ~. \tag{5.5d}$$

We use the metric $(\eta_{ab}) = \text{diag.}(+,-,\cdots,-)$, and exactly the same notation as in [18], whose details we skip here. As before, other independent components, such as $H_{\alpha b_1\cdots b_{11}}$ are all zero.

In the confirmation of BIs at $d \leq 1$, some crucial relations are needed. For example, the $(\alpha\beta\gamma\delta e_1\cdots e_9)$-type $H$-BI at $d=0$ needs:

$$(\gamma^f)_{(\alpha\beta|}(\gamma_{fe_1\cdots e_9})_{|\gamma\delta)} = (\gamma^f)_{(\alpha\beta|}(i\epsilon_{fe_1\cdots e_9}{}^g \gamma_g)_{|\gamma\delta)} = +i\epsilon_{e_1\cdots e_9}{}^{fg}(\gamma_f)_{(\alpha\beta|}(\gamma_g)_{|\gamma\delta)} \equiv 0 ~. \tag{5.6}$$

Another example is the $(\alpha\beta c_1\cdots c_{11})$-type $H$-BI at $d=1$ requiring

$$(\gamma_{[c_1\cdots c_8|}{}^d)_{\alpha\beta} F_{d|c_9c_{10}c_{11}]} \equiv 0 ~, \tag{5.7}$$

which is confirmed by its multiplication by $\epsilon^{c_1\cdots c_{11}}$:

$$\epsilon^{c_1\cdots c_{11}}(\gamma_{c_1\cdots c_8}{}^d)_{\alpha\beta} F_{dc_9c_{10}c_{11}} = -\tfrac{i}{2}\epsilon^{c_1\cdots c_{11}}\epsilon_{c_1\cdots c_8}{}^{def}(\gamma_{ef})_{\alpha\beta} F_{dc_9c_{10}c_{11}} \equiv 0 ~, \tag{5.8}$$

because one of the indices $c_9$, $c_{10}$, $c_{11}$ on $F_{dc_9c_{10}c_{11}}$ should be $d$, like eq. (2.13).

We have now seen that we can introduce extra superfield strengths even for 11D supergravity, contrary to common wisdom about its tight field content. However, we also add that our modification alters only the off-shell structure, with no essential couplings between $N$'s and the original physical fields, up to topological considerations. This is also expected from the past trials of modifying the 11D supergravity [19]. Note that there is an important difference of the 11D case from the massive supergravity in 10D of the previous section. In 10D the superfield strength $N_{A_1 \cdots A_{10}}$ is involved non-trivially in constraint such as (2.10n), while in 11D $N_{A_1 \cdots A_{11}}$ has no such non-trivial couplings with physical fields. Relevantly, as contrast to the 10D case, the cosmological constant vanishes in the 11D case.

We can further try to repeat the construction of super tenbrane action, following the super ninebrane result in 10D. However, we soon notice an obstruction that the 11D analog of $\Gamma$-matrix (4.2b) is reduced to an identity, due to the 11D $\gamma$-matrix relation $\epsilon^{a_1 \cdots a_{11}} \gamma_{a_1 \cdots a_{11}} = +i(11!)I$.

## 6. Super $(2k-1)$-Brane Action in $D = 2k$

Since we have seen how the super ninebrane formulation works in 10D, it is straightforward to generalize it to an arbitrary supergravity theory in even space-time dimensions $D = 2k$. The starting point is to establish the superspace formulation with the $D$-form and $(D+1)$-form superfield strengths $N_{A_1 \cdots A_D}$ and $H_{A_1 \cdots A_{D+1}}$, with the BIs:

$$\tfrac{1}{D!} \nabla_{[A_1} N_{A_2 \cdots A_{D+1})} - \tfrac{1}{2[(D-1)!]} T_{[A_1 A_2|}{}^B N_{B|A_3 \cdots A_{D+1})} - H_{A_1 \cdots A_{D+1}} N \equiv 0 \ , \tag{6.1}$$

$$\tfrac{1}{(D+1)!} \nabla_{[A_1} H_{A_2 \cdots A_{D+2})} - \tfrac{1}{2(D!)} T_{[A_1 A_2|}{}^B H_{B|A_3 \cdots A_{D+2})} \equiv 0 \ , \tag{6.2}$$

together with other conventional BIs in the original $D$-dimensional supergravity. Now these new BIs should be satisfied by the constraints

$$T_{\underline{\alpha\beta}}{}^c = i(\gamma^c)_{\underline{\alpha\beta}} \ , \quad \cdots \tag{6.3a}$$

$$N \equiv \tfrac{1}{D!} \epsilon^{a_1 \cdots a_D} N_{a_1 \cdots a_D} = \text{const.} \ , \quad \nabla_A N = 0 \ , \tag{6.3b}$$

$$H_{\underline{\alpha\beta} c_1 \cdots c_{D-1}} = \begin{cases} i \left( \gamma_{c_1 \cdots c_{D-1}} \gamma_{D+1} \right)_{\underline{\alpha\beta}} & (\text{for } D = \text{ even}) \ , \\ \left( \gamma_{c_1 \cdots c_{D-1}} \right)_{\underline{\alpha\beta}} & (\text{for } D = \text{ odd}) \ , \end{cases} \tag{6.3c}$$

where $\cdots$ in (6.3a) denote other necessary constraints for the original $D$-dimensional supergravity itself, whose details depend on the dimensions. Also depending on $D$ is the dottedness of the $\gamma$-matrices, which is implicitly included in the underlined spinorial indices here. These underlined indices also include any possible $N \geq 2$ indices in some dimensions such as $D = 6$ [20], where $(\gamma^c)_{\alpha\beta}$ are antisymmetric, so that additional $N \geq 2$ indices $i, j, \cdots$ are needed, as $T_{\underline{\alpha\beta}}{}^c = i(\gamma^c)_{\alpha\beta} \epsilon_{ij}$ with an antisymmetric metric $\epsilon_{ij}$. Additionally, because of the contraction rules for spinorial indices, with or without antisymmetric charge-conjugation matrices depending on $D$, equations in (6.3) (and also in (6.8) below) are correct up to signatures. As before, all other independent components of $N$ or $H$-superfield strengths, such as $H_{\underline{\alpha} b_1 \cdots b_D}$ are zero.

It is straightforward to show how all the BIs (6.1) and (6.2) at $d = 0$ can be satisfied by (6.3). The most crucial identity is for $H$-BI at $d = 0$: When $D = $ odd, the identity is

$$I_{\underline{\alpha\beta\gamma\delta}} \equiv (\gamma_a)_{\underline{\alpha\beta}} \left( \gamma^{ab_1 \cdots b_{D-2}} \right)_{\underline{\gamma\delta}} \underset{10}{=} c \epsilon^{ab_1 \cdots b_{D-2} c} (\gamma_a)_{\underline{\alpha\beta}} (\gamma_c)_{\underline{\gamma\delta}} \equiv 0 \ , \tag{6.4}$$

with a constant $c$. This is because $\gamma^{[D-1]}$ is $\epsilon$-tensor times one $\gamma$-matrix in odd dimensions. While if $D =$ even, then the identity is

$$I_{\underline{\alpha\beta\gamma\delta}} \equiv (\gamma_a)_{\underline{\alpha\beta}} (\gamma^{ab_1\cdots b_{D-2}}\gamma_{D+1})_{\underline{\gamma\delta}} = c'\epsilon^{ab_1\cdots b_{D-2}c} (\gamma_a)_{\underline{\alpha\beta}} (\gamma_c)_{\underline{\gamma\delta}} \equiv 0 \ , \tag{6.5}$$

for $\gamma_{D+1}$ proportional to $\gamma_0\gamma_1\cdots\gamma_{D-1}$. Thus whether $D$ is odd or even, the $H$-BI at $d=0$ is identically satisfied, due to the total antisymmetry of the $\epsilon$-tensor.

In order to satisfy $d=1/2$ and $d=1$ $H$-BIs, we need two conditions, respectively

$$T_{(\underline{\alpha\beta}|}{}^{\underline{\epsilon}}H_{\underline{\epsilon}|\underline{\gamma})d_1\cdots d_{D-1}} = 0 \ , \tag{6.6}$$

$$T_{[c_1|(\underline{\alpha}|}{}^{\underline{\delta}}H_{\underline{\delta}|\underline{\beta})|c_2\cdots c_D]} = 0 \ . \tag{6.7}$$

In fact, our previous 11D case satisfies both of these: (6.6) trivially, and (6.7) by the algebraic identity (5.8). Since the explicit structure of the components $T_{\underline{\alpha\beta}}{}^{\underline{\gamma}}$ and $T_{\underline{\alpha}b}{}^{\underline{\gamma}}$ depends on the supergravity theory in $D$-dimensions, we do not go into the details any more. As for the $N$-BIs, one at $d=1/2$ is rather trivial, while the old problem at $d=1$ does not arise, thanks to the $HN$-term in (6.2).

If this superspace formulation up to now is the whole story without a super $p$-brane action, the space-time dimensions can be both even and odd. However, as we have seen for 11D case, due to the triviality of the matrix $\Gamma$ in $D=$ odd, only $D=$ even is allowed for a super $(D-1)$-brane action with the fermionic $\kappa$-symmetry. Considering this, let us give the ansatz for the super $(2k-1)$-brane action for $D = 2k \leq 10$, as a generalization of (4.1) - (4.5), or even as a special case of super $p$-brane [2]: Our relevant equations are[7]

$$I_{2k} \equiv \int d^{2k}\sigma \left[ +\tfrac{1}{2}\sqrt{-g}g^{ij}\Pi_i{}^a\Pi_{ja} - \tfrac{D-2}{2}\sqrt{-g} \right.$$
$$\left. + \tfrac{1}{(2k)!}\epsilon^{i_1\cdots i_{2k}}\Pi_{i_1}{}^{A_1}\cdots\Pi_{i_{2k}}{}^{A_{2k}}C_{A_{2k}\cdots A_1} \right] \ , \tag{6.8a}$$

$$\delta_\kappa E^{\underline{\alpha}} = \tfrac{1}{2}(I+\Gamma)^{\underline{\alpha}}{}_{\underline{\beta}}\kappa^{\underline{\beta}} \ , \quad \delta_\kappa E^a = 0 \ , \tag{6.8b}$$

$$\Gamma^{\underline{\alpha}}{}_{\underline{\beta}} \equiv \tfrac{(-1)^{(k-1)(2k+1)/2}}{(2k)!\sqrt{-g}}\epsilon^{i_1\cdots i_{2k}}\Pi_{i_1}{}^{a_1}\cdots\Pi_{i_{2k}}{}^{a_{2k}}(\gamma_{a_1\cdots a_{2k}})^{\underline{\alpha}}{}_{\underline{\beta}} = (\gamma_{2k+1})^{\underline{\alpha}}{}_{\underline{\beta}} \ , \tag{6.8c}$$

$$g_{ij} = \Pi_i{}^a\Pi_{ja} \ . \tag{6.8d}$$

The last equality in (6.8c) is the crucial one, in our case with the same dimensionality both for the super-worldvolume and the target space-time, under the embedding equation (6.8d).

Once we have seen the existence of over-ranked superfield strengths in 11D, 10D and

---

[7]We use $2k$ for the dimensionality, not using the word 'super D-brane' in $D$-dimensions to avoid the confusion with Dirichlet $p$-brane [3].

$D \leq 9$, we notice that there is an interesting sequence among them depicted by the diagram:

$$\begin{array}{ccccc} \mathbf{D=11} & \longrightarrow & \mathbf{D=10} & \longrightarrow & \cdots \\ H_{A_1 \cdots A_{12}} & & & & \\ N_{A_1 \cdots A_{11}} & \longrightarrow & H_{A_1 \cdots A_{11}} & & \\ & & N_{A_1 \cdots A_{10}} & \longrightarrow & \cdots \end{array} \quad (6.9)$$

connected by appropriate dimensional reductions. The essential point is that the 11-form $N_{A_1 \cdots A_{11}}$ in 11D can be the origin of the over-ranked 11-form $H_{A_1 \cdots A_{11}}$ in 10D, and the same pattern seems to continue to lower dimensions $D \leq 9$. Since the M-theory is the underlying non-perturbative theory in 11D [21][22], this result may well provide a new important link or duality between M-theory and 10D superstring, or even with theories in $D \geq 12$ [23][24] which in turn provides the origin of $H_{A_1 \cdots A_{12}}$ in 11D itself.

## 7. Concluding Remarks

In this paper, we have presented a superspace formulation with the 10-form superfield strength $N_{A_1 \cdots A_{10}}$, as the most important foundation for super eightbrane theory [1] that has not been performed in the past. We have found a remarkable fact that the over-ranked 11-form superfield strength $H_{A_1 \cdots A_{11}}$ is the crucial key for the $N$-BI to be satisfied. The peculiar feature of our superspace formulation is triple-fold: First, the rank 11 of $H_{A_1 \cdots A_{11}}$ exceeds the space-time dimension 10. Second, $H_{A_1 \cdots A_{11}}$ appears in the superfield strength $N_{A_1 \cdots A_{10}}$ as a generalized Chern-Simons term, associated with the $HN$-term in the $N$-BI (2.4). Third, the whole mechanism works, only when the scalar superfield $N$ is constant: $\nabla_{\underline{\alpha}} N = 0$. To our knowledge, there has been so far no such a superspace formulation with an over-ranked superfield strength involved in a peculiar generalized Chern-Simons term. Taking also the advantage of the simplest structure of the BFFC constraints [12], we have drastically simplified our superspace computation.

We have also presented a super ninebrane action with 10D world-supervolume with the WZNW term, naturally expected from the presence of the 10-form potential superfield $C_{A_1 \cdots A_{10}}$. It seems now that the super ninebrane action is a natural result of super eightbrane formulation itself. In the conventional $p$-brane context [2], there was no strong motivation for over-ranked superfield strengths. Especially, the possibility of super ninebrane formulation in 10D, in which the dimensionality of the target space-time coincides that of the super-worldvolume, was not seriously considered, ever since the first super $p$-brane formulation [2]. The recent development of super eightbrane [1][2] or Dirichlet eightbrane [3] gave a strong motivation of introducing such over-ranked superfield strengths. We also mention that our ninebrane action with the $\kappa$-symmetry may be regarded as a non-linear realization of supersymmetry in 10D, similarly to the Dirichlet brane action in [25].

As a by-product of our 10D result with an over-ranked superfield strength, we have also confirmed that a 12-form superfield strength can be introduced also into 11D superspace. As for the possibility of super tenbrane formulation in 11D, there seems to be an obstruction about the matrix $\Gamma$ [2], because its usual definition results in an identity matrix. Despite of this situation in 11D, our result for over-ranked superfield strength seems to have lots of applications even to $D \leq 9$. Namely, whenever there is a 'massive' supergravity in $D$-dimensions, we first introduce the $D$-form field strength $N_{a_1 \cdots a_D}$ as a dual to a scalar field $N$ replacing the mass parameter $m$. We then introduce an over-ranked superfield strength $H_{A_1 \cdots A_{D+1}}$ in order to satisfy the $N$-BIs by a generalized Chern-Simons term. However, a super $(D-1)$-brane formulation does not seem to exist in $D$-dimensions for $D = $ odd, due to the property of Clifford algebras forbidding as in 11D the non-trivial matrix $\Gamma$ needed for a $\kappa$-symmetry.

Another consequence of our result is that the apparently 'old-fashioned' superspace approach [6] is still powerful to discover yet unknown features of super $p$-brane physics. This is because we can not even construct super $p$-brane action [2] without establishing the underlying superspace for backgrounds. Accordingly, there seems to be no alternative way other than introducing the over-ranked superfield strength $H_{A_1 \cdots A_{11}}$. Moreover, such an over-ranked field strength does not enter the algebraic analysis of supersymmetries for M-theory in terms of Clifford algebra [22]. ¿From these viewpoints, superspace formulation [6] still maintains its usefulness for finding unexpected features in supergeometry, even nowadays. There seems no alternative quick way to avoid the struggles with tremendous amount of the $\gamma$-matrix algebra, as we have performed for the satisfaction of our BIs.

Our result is also suggestive of possible underlying supergravity theory in $D \geq 12$ like those in [24], as the origin of the over-ranked superfield strength in $10D$. It is plausible that higher-dimensional non-perturbative theories [23] which are supposed to underlie the 10D or 11D theories, may well enter the configuration, when we start considering super $p$-brane physics as the basis of non-perturbative description of superstrings or supermembranes.

Considering that there has been no superspace formulation for super eightbrane or Dirichlet eightbrane in the past, we re-stress the importance of our result as the establishment of supergeometry for super eightbrane [1] for the first time. It is natural that our formulation, due to its unconventional over-ranked superfield strength involved in a peculiar Chern-Simons term, has been overlooked for such a long time since the first component formulation effective theory of super eightbrane [1]. We expect more future developments related to our new superspace formulation with over-ranked superfield strengths in higher $(D \geq 12)$ [24] as well as lower $(D \leq 9)$ dimensional theories.

We are grateful to S.J. Gates, Jr. for various helpful discussions, that led to the major breakthroughs in this paper. We are also indebted to E. Bergshoeff, M.B. Green and P.K. Townsend for information about ref. [1]. Additionally, we are indebted to the referee of this paper for various suggestions to improve the paper.